\documentclass[12pt]{iopart}

\usepackage{graphicx}
\usepackage{color}
\usepackage{units}
\usepackage{psfrag}
\usepackage{multirow}
\usepackage{tabularx}

\begin{document}

\title[]{Relaxation and fluctuation dynamics in coherent two-dimensional electronic spectra}

\author{Joachim Seibt, T\~{o}nu Pullerits}

\address{Department of Chemical Physics, Lund University, Box 124, SE-2100, Lund, Sweden}

\ead{Joachim.Seibt@chemphys.lu.se, Tonu.Pullerits@chemphys.lu.se}

\begin{abstract}
Two-dimensional (2D) spectroscopy provides information about dissipative processes subsequent to electronic excitation, which play a functional role in energy harvesting materials and devices. This technique is particularly sensitive to electronic and vibronic coherence dynamics.
While the theoretical treatment of relaxation in the context of 2D-spectroscopy is well-developed under the assumption of different timescales of population transfer and fluctuation dynamics, the interplay between both kinds of processes lacks a comprehensive description in terms of line shape functions.
To bridge this gap, we use the cumulant expansion approach to derive response functions, which account for fluctuation dynamics and population transfer simultaneously. We compare 2D-spectra from calculations with different model assumptions about correlations between fluctuations and point out under which conditions a simplified treatment is justified.
Our study shows that population transfer and dissipative fluctuation dynamics cannot be described independent of each other in general. Advantages and limitations of the proposed calculation method and its compatibility with the modified Redfield description are discussed.

\end{abstract}

\maketitle

\section{Introduction} \label{sec:Introduction}

Relaxation and dephasing dynamics in molecular aggregates and nanoparticles subsequent to electronic excitation can be investigated in detail via two-dimensional (2D) spectroscopy.  
Compared to pump-probe spectroscopy, this technique has the advantage that it allows to reveal information about coherences without limitations of the spectral resolution determined by the pulse width \cite{Brixner05NAT}.
The theoretical and experimental aspects of 2D-spectroscopy are extensively described in the literature \cite{BrMaSt04_4221_,Faeder99JCP,Mukamel04CR,ChoCR2008}. 
Specific ways of calculating 2D-spectra, such as a non-perturbative approach \cite{Seidner95JCP,Mancal06JCP} or description of the nonlinear response on the Keldysh contour \cite{HaPu12_JPB} were proposed.
Previous theoretical investigations in the context of 2D electronic spectroscopy include the application to quantum dots \cite{SeHaPu13_JPCB,SePu13_JPCC_18728,TuHaSc12_NL_880}, dimer systems \cite{BuVaAb14_JCP,Kjellberg06PhysRevB,Chen09JCP,SeEi12_JCP_024109,SeReVo09_134318_} and light harvesting complexes \cite{BrKjPu07_CPL_192,Sharp10JCP,HeKrKr12_NJP_023018}, where for the latter, besides third-order processes, also fifth-order processes have been described \cite{BrPu11_NJP_025024}.
Recently, the origin of long coherence lifetime in the FMO complex was discussed.
These coherences had been interpreted initially as a purely electronic effect \cite{PaVoAb11_PNAS_20908}. Following different lines of argumentation, the role of vibrational \cite{TiPeJo13_PNAS_1203} and vibronic effects \cite{ChKaPuMa12_JPCB_7449,PoKuPu_CP_21,ScToKu14_NJP_045010} for the persistence of the coherence beatings reflected by crosspeak oscillations in 2D-spectra was pointed out \cite{PuZiSu13_PNAS_1148}. Signatures of vibrational and electronic coherence beatings in 2D-spectra in terms of amplitude and phase relationships of the oscillations were identified \cite{BuZiVaAb12_CPL_40}, and an approach enabling quantification of vibrational effects in coherence beatings of molecular aggregates was introduced \cite{ChChKa13_SR}. Furthermore, the interplay between electronic and vibrational degrees of freedom was reported as a key aspect for the long coherence lifetime \cite{PlAmHu13_JCP_235102}.

For the description of dissipation under the influence of an environment, different approaches can be used \cite{May11}, including a treatment in terms of surface hopping \cite{BeDaKj02_JCP_5810} and stochastic Schr\"{o}dinger equations \cite{ScFaKe14_JPCA}.
In the context of response functions, usually either density matrix propagation \cite{Egorova07JCP} or a line shape function approach \cite{BuVaAb12_JCP_044513} is chosen.
In the first case, the excitation-induced relaxation dynamics under the influence of environment fluctuations enters in the Liouville-von-Neumann equation by including a dissipative term.
Under the assumption of a Markovian environment the relaxation process can be described by using the Redfield approach \cite{PiMaFl06_234505_,Kjellberg05JCP} or the related Lindblad formulation \cite{PlAmHu13_JCP_235102}.
In the latter case, dissipation is taken into account in terms of expressions containing line shape functions, which are obtained from a second order cumulant expansion \cite{Mukamel95}. Following this approach, population decay can be easily included, while population transfer is often neglected \cite{SeHaPu13_JPCB} or taken into account using a simplified phenomenological rate equation treatment \cite{ChPoYaPu09_PRB_245118}.
Combinations of a density matrix description of relaxation between electronic states and treatment of fluctuations in terms of line shape functions have also been proposed \cite{AbMu11_JCP_174504,BuVaAb14_JCP}, however without taking the interplay between both types of processes into account.
Previous descriptions of the regime, where population transfer and fluctuation timescales are comparable and the two processes influence each other, have been formulated in terms of stochastic Liouville equations \cite{Abramavicius09CR,Sanda08JPCB}.
For an approximate treatment of the interaction of the system with the environment, including non-Markovian effects, different methods \cite{Tanimura06JPCJ,RoEiWoSt09_PRL_058301,NaIsFl11_NJP_063040} exist, where the HEOM approach \cite{Tanimura06JPCJ} has been used in the context of 2D spectroscopy \cite{Chen09JCP,HeKrKr12_NJP_023018}.

In this work the concept of rate equations is combined with the line shape function approach. By using the second-order cumulant expansion, we derive rigorous general expressions for the response functions of a system with two singly excited sub-levels, between which population transfer can appear subsequent to optical excitation.
The derivation leads to self-contained third-order response functions including the interplay between fluctuations and population transfer, which is usually neglected.
A similar approach, however related to a more specific system consisting of a donor and an acceptor molecule, has been discussed previously \cite{Yang99JCP,MaDoPs14_CJC_135}.
As in these works, also in our investigation the population transfer rates are assumed to be independent of the fluctuations of the environment.
In general, the rate constants are determined by fluctuations, as described by the modified Redfield theory \cite{ScKlSc06_JCP_,YaFl_02_CP_163,ZhMeCh98_JCP_7763}, which, in principle, can be combined with our approach.
Such a combined description of the dynamics, including time-dependent population transfer rates, is only outlined in this work. 
If this way of treatment is chosen, the dynamics are completely governed by line shape functions, which contain all orders of the system-bath interaction \cite{OlToCh13_arXiv}.
While the comprehensive treatment of the system-bath interaction in terms of line shape functions yields a benchmark for the approximative HEOM approach, the modified Redfield description of population dynamics and lifetime broadening of coherences does not include the non-Markovian effects which enter in HEOM \cite{HsStSe10_NJP_085005}.

The article is organized as follows: 
First we present the theoretical background containing a derivation of the expressions for a combined description of fluctuation and relaxation dynamics by using the cumulant expansion. Furthermore, the concrete way of calculating 2D spectra of our model system is described.
In the Results part we investigate how the 2D-spectra are influenced by the relaxation dynamics and the model assumptions about correlations between fluctuations related to different singly excited sublevels. We figure out under which conditions the population transfer and the dissipative fluctuation dynamics can be considered separately. This study also includes a comparison of our general approach with results for a long time limit, which has been addressed previously in the literature.
Finally, we discuss the appearance of a phase shift between vibrational beatings in the population transfer crosspeak evolution under the assumptions of correlated and uncorrelated fluctuations.  

\section{Theoretical background}

\subsection{Derivation of response functions including population transfer by using the cumulant expansion}

Third-order system-field interaction processes in the optical spectral range can be described in a perturbative way by using response functions, which depend on the time delays $t_1$, $t_2$ and $t_3$ between the interaction instances.
In the following, population transfer processes and fluctuation dynamics during the population time $t_2$ are taken into account explicitly.
In the framework of the secular approximation, where populations and coherences do not mix, the response functions related to the stimulated emission (SE) and excited state absorption (ESA) processes can be separated in terms of components with coherence and population evolution in the singly excited state during $t_2$ \cite{SeEi12_JCP_024109}.
In total there a ten different such response functions.
In the following we derive the expression for the SE-type response function $R_{2g}$ \cite{BrMaSt04_4221_} with population transfer between singly excited states $\alpha$ and $\beta$ after excitation from the electronic ground state $g$. The derivation of all others follows the same steps and is not shown. All final expressions are provided in the Appendix.
The derivation starts from a Liouville space formulation \cite{Mukamel95}, which contains matrix elements of the Green operator ${\cal G}$, the dipole operator $V$ and its Liouville space analogue ${\cal V}$. Matrix elements of the operator ${\cal V}$ in the basis of the electronic states account for the influence of a system-field interaction term in the Liouville-von-Neumann equation and describe instantaneous transitions between the involved electronic states. Initially only the electronic ground state is populated, i.e.\ the only non-vanishing element of the initial density matrix $\rho_g$ appears at the diagonal position related to $g$.
Population transfer between the singly excited states can be taken into account in terms of tensor elements of the relaxation superoperator ${\cal K}$ in Liouville space. 
Relaxation between $\alpha$ and $\beta$ is considered as a Markovian process, which is reflected by the assumption of a time-local transfer event from $\alpha$ to $\beta$ at time $s$ selected from the interval between $0$ to $t_2$.
The continous relaxation dynamics can be taken into account by convolution with a $s$-dependent function for the description of population evolution.

As a preparative step for the cumulant expansion, the initial Liouville space formulation of the third-order response is transformed into a Hilbert space representation \cite{Mukamel95}. To this end, the matrix elements of the Liouville space operators are expressed in terms of commutators $[\bullet,\bullet]_{-}$ (following the ``$\bullet$'' notation for the action of operators from \cite{May11}) as

\begin{eqnarray} \label{eq:interaction_operator_commutators}
&&{\cal G}^{\dagger}_{\kappa \lambda, \kappa \lambda}(t') {\cal V}_{\kappa \lambda, \mu \nu} {\cal G}_{\mu \nu, \mu \nu}(t') \; \bullet 
= \delta_{\lambda \nu}[V_{\kappa \mu}(t'), \bullet]_{-} - \delta_{\kappa \mu}[V_{\nu \lambda}(t'), \bullet]_{-},
\nonumber \\ 
&&\bullet \; {\cal G}^{\dagger}_{\kappa \lambda, \kappa \lambda}(t') {\cal V}_{\kappa \lambda, \mu \nu} {\cal G}_{\mu \nu, \mu \nu}(t')   
= \delta_{\lambda \nu}[\bullet,V_{\kappa \mu}(t')]_{-} - \delta_{\kappa \mu}[\bullet,V_{\nu \lambda}(t')]_{-},
\nonumber \\ 
&&\{ \kappa, \lambda, \mu, \nu \} \in \{ \alpha, \beta, g \}.
\end{eqnarray}

The relaxation superoperator for population transfer from $\alpha$ to $\beta$ is formulated in terms of matrix elements of the system-bath coupling operator $\Theta$ within the framework of the Redfield approach \cite{May11} as

\begin{eqnarray} \label{eq:relaxation_tensor_commutators}
&&{\cal G}^{\dagger}_{\beta \beta, \beta \beta}(t') {\cal K}_{\beta \beta, \alpha \alpha} {\cal G}_{\alpha \alpha, \alpha \alpha}(t') \; \bullet = [\Theta_{\alpha \beta}(t'),\Theta_{\beta \alpha}(t') \; \bullet - \bullet \; \Theta_{\alpha \beta}(t')]_{-},
\nonumber \\ 
&&\bullet \; {\cal G}^{\dagger}_{\beta \beta, \beta \beta}(t') {\cal K}_{\beta \beta, \alpha \alpha} {\cal G}_{\alpha \alpha, \alpha \alpha}(t') = [ \bullet \; \Theta_{\beta \alpha}(t') - \Theta_{\alpha \beta}(t') \; \bullet,\Theta_{\alpha \beta}(t')]_{-}
\end{eqnarray}

Matrix elements of $\Theta$ account for transitions between the electronic states given by the indices. The Redfield description includes the assumption that these transitions are facilitated by bath phonons.

While the initial formula in Liouville space in analogy to \cite{Sanda08JPCB} can be written as

\begin{eqnarray}
R_{2g,\alpha \beta}(t_1,t_2,t_3,s) &=& \langle \langle V_{\beta g} | {\cal G}_{\beta g, \beta g}(t_3) {\cal V}_{\beta g, \beta \beta} {\cal G}_{\beta \beta, \beta \beta}(t_2-s) {\cal K}_{\beta \beta, \alpha \alpha} 
\nonumber 
\\
&&{\cal G}_{\alpha \alpha, \alpha \alpha}(s) {\cal V}_{\alpha \alpha, g \alpha} {\cal G}_{g \alpha, g \alpha}(t_1) {\cal V}_{g \alpha, g g} | \rho_g \rangle \rangle,
\end{eqnarray}

it becomes
 
\begin{eqnarray} \label{eq:response_function_commutators_Theta}
R_{2g,\alpha \beta}(t_1,t_2,t_3,s)&=&
-\langle [[[\Theta_{\alpha \beta}(t_1+s) [V_{g \beta}(t_1+t_2+t_3),V_{\beta g}(t_1+t_2)]_{-},
\nonumber
\\
&&\Theta_{\alpha \beta}(t_1+s)]_{-},V_{\alpha g}(t_1)]_{-},V_{g \alpha}(0)]_{-}\rho_{g}\rangle
\nonumber
\\
&&+\langle [[[V_{g \beta}(t_1+t_2+t_3),V_{\beta g}(t_1+t_2)]_{-} \Theta_{\beta \alpha}(t_1+s),
\nonumber
\\
&&\Theta_{\alpha \beta}(t_1+s)]_{-},V_{\alpha g}(t_1)]_{-}, V_{g \alpha}(0)]_{-}\rho_{g}\rangle.
\end{eqnarray} 
 
in Hilbert space.
Vanishing terms in the expansion of the commutator expressions can be determined by checking whether adjacent indices of neighbored operators are different. 
To obtain a result different from zero, equal indices $\lambda$ and $\mu$ in $O_{\kappa \lambda} O_{\mu \nu}$ with $O \in \{ V, \Theta \}$ are required.
Furthermore, for a non-vanishing trace the first index of the operator at the leftmost position of $< O_{\kappa \lambda} ... O_{\mu \nu} \rho_{\nu}>$ needs to fulfill the condition $\kappa=\nu$, i.e.\ $\kappa=g$ in our case.

These considerations lead to a single remaining term

\begin{eqnarray} \label{eq:remaining_term}
R_{2g,\alpha \beta}(t_1,t_2,t_3,s) &=& \langle V_{g \alpha}(0) \Theta_{\alpha \beta}(t_1+s) V_{\beta g}(t_1+t_2) V_{g \beta}(t_1+t_2+t_3) 
\nonumber
\\
&&\Theta_{\beta \alpha}(t_1+s) V_{\alpha g}(t_1) \rho_{g} \rangle.
\end{eqnarray}

Before the cumulant expansion is performed, $\Theta_{\alpha \beta}(t_1+s)$ and $\Theta_{\beta \alpha}(t_1+s)$ are expressed via $\Theta_{\alpha g}(t_1+s)$,
$\Theta_{g \beta}(t_1+s)$, $\Theta_{\beta g}(t_1+s)$ and $\Theta_{g \alpha}(t_1+s)$, as the electronic ground state is taken as a reference state. 
Expressing the time evolution operators in terms of exponentials with different time-ordering, indicated as $\exp_{+}$ and $\exp_{-}$ \cite{Mukamel95}, leads to

\begin{eqnarray}
\Theta_{\alpha \beta}(t')&=&\exp(i H_{\alpha} t') \Theta_{\alpha \beta} \exp(-i H_{\beta} t')
\nonumber
\\
&=&\exp(i H_{\alpha} t')\exp(-i H_{g} t') \Theta_{\alpha \beta} \exp(i H_{g} t')\exp(-i H_{\beta} t')
\nonumber
\\
&=&\exp(i \omega_{\alpha g} t')\exp_{-}\left[i \int_{0}^{t'} dt'' U_{\alpha}(t'')\right] \Theta_{\alpha \beta} 
\nonumber
\\
&&\exp(-i \omega_{\beta g} t')\exp_{+}\left[-i \int_{0}^{t'} dt'' U_{\beta}(t'')\right]
\end{eqnarray}

and

\begin{eqnarray}
\Theta_{\beta \alpha}(t')&=&\exp(i H_{\beta} t') \Theta_{\beta \alpha} \exp(-i H_{\alpha} t')
\nonumber
\\
&=&\exp(i H_{\beta} t')\exp(-i H_{g} t') \Theta_{\beta \alpha} \exp(i H_{g} t')\exp(-i H_{\alpha} t')
\nonumber
\\
&=&\exp(i \omega_{\beta g} t')\exp_{-}\left[i \int_{0}^{t'} dt'' U_{\beta}(t'')\right] \Theta_{\beta \alpha} 
\nonumber
\\
&&\exp(-i \omega_{\alpha g} t')\exp_{+}\left[-i \int_{0}^{t'} dt'' U_{\alpha}(t'')\right]
\end{eqnarray}

After setting all time-independent matrix elements of $V$ and $\theta$ equal to $1$ and introducing time-ordered exponentials also for the transition dipole operator matrix elements, Eq.~(\ref{eq:remaining_term}) becomes

\begin{eqnarray} \label{eq:remaining_term_time_evolution}
R_{2g,\alpha \beta}(t_1,t_2,t_3,s) &=& \exp(i \omega_{\alpha g} t_1) \exp(-i \omega_{\beta g} t_3) 
\nonumber
\\
&&\left<
\exp_{+}\left[-i \int_{0}^{0} d\tau_1 U_{\alpha}(\tau_1)\right] \exp_{-}\left[i \int_{0}^{t_1+s} d\tau_2 U_{\alpha}(\tau_2)\right] \right.
\nonumber
\\
&&\exp_{+}\left[-i \int_{0}^{t_1+s} d\tau_3 U_{\beta}(\tau_3)\right] \exp_{-}\left[i \int_{0}^{t_1+t_2} d\tau_4 U_{\beta}(\tau_4)\right]
\nonumber
\\
&&\exp_{+}\left[-i \int_{0}^{t_1+t_2+t_3} d\tau_5 U_{\beta}(\tau_5)\right] \exp_{-}\left[i \int_{0}^{t_1+s} d\tau_6 U_{\beta}(\tau_6)\right]
\nonumber
\\
&&\left. \exp_{+}\left[-i \int_{0}^{t_1+s} d\tau_7 U_{\alpha}(\tau_7)\right] \exp_{-}\left[i \int_{0}^{t_1} d\tau_8 U_{\alpha}(\tau_8)\right]
\right>.
\end{eqnarray}

Second-order cumulant expansion of Eq.~(\ref{eq:remaining_term_time_evolution}) \cite{Mukamel95} and formulation of the resulting expression in terms of line shape functions $g_{i j}$ related to the singly excited states $\{i,j\} \in \{\alpha,\beta\}$ leads to the population transfer term

\begin{eqnarray} \label{eq:remaining_term_line_shape_functions}
R_{2g,\alpha \beta}(t_1,t_2,t_3,s) &=& \exp(i \omega_{\alpha g} t_1) \exp(-i \omega_{\beta g} t_3) 
\nonumber
\\
&&\exp \left( 
-g^{*}_{\alpha \alpha}(t_1)+g_{\alpha \beta}(t_2)-g^{*}_{\beta \beta}(t_3) \right.
\nonumber
\\
&&\left. -g^{*}_{\alpha \beta}(t_1+t_2)-g_{\alpha \beta}(t_2+t_3)+g^{*}_{\alpha \beta}(t_1+t_2+t_3) \right.
\nonumber
\\
&&\left.
+2 i \Im(g_{\beta \beta}(t_2-s))-2 i \Im(g_{\alpha \beta}(t_2-s)) \right.
\nonumber
\\
&&\left. +2 i \Im(g_{\alpha \beta}(t_2-s+t_3))-2 i \Im(g_{\beta \beta}(t_2-s+t_3))
\right).
\end{eqnarray}

The imaginary parts of the line shape function components with dependence on the variable $s$, which enter with prefactors $2i$ in the argument of the exponential, lead to a frequency shift, which varies as a function of $s$. 
Note that the population dynamics is not included in the $s$-dependent expressions $R_{2g,\alpha \alpha}(t_1,t_2,t_3,s)$, $R_{2g,\alpha \beta}(t_1,t_2,t_3,s)$, $R_{2g,\beta \alpha}(t_1,t_2,t_3,s)$ and $R_{2g,\beta \beta}(t_1,t_2,t_3,s)$ yet. 

Under the assumption of phenomenological rate constants $\Gamma_{\alpha \beta}$ for population transfer from $\alpha$ to $\beta$ and $\Gamma_{\beta \alpha}$ for population transfer from $\beta$ to $\alpha$, the population dynamics can be expressed by the relaxation tensor \cite{SeEi12_JCP_024109}

\begin{equation} \label{eq:relaxation_dynamics}
G_{k k l l} \left( t_2 \right) = 
\left[
\left(
\begin{array}{cccc}
1 & 0 & 0 & 0 \\
0 & \frac{\Gamma_{\alpha \beta}f(t_2)+\Gamma_{\beta \alpha}}{\Gamma_{\alpha \beta}+\Gamma_{\beta \alpha}} & -\frac{\Gamma_{\beta \alpha}(f(t_2)-1)}{\Gamma_{\alpha \beta}+\Gamma_{\beta \alpha}} & 0 \\
0 & -\frac{\Gamma_{\alpha \beta}(f(t_2)-1)}{\Gamma_{\alpha \beta}+\Gamma_{\beta \alpha}} & \frac{\Gamma_{\beta \alpha}f(t_2)+\Gamma_{\alpha \beta}}{\Gamma_{\alpha \beta}+\Gamma_{\beta \alpha}} & 0 \\
0 & 0 & 0 & 1 
\end{array}
\right)
\right]_{k l},
\end{equation}

with $f(t_2)=\exp(-(\Gamma_{\alpha \beta}+\Gamma_{\beta \alpha})t_2)$.
By convolution of the derived $s$-dependent line shape function expressions with the component-wise differential of the relaxation tensor at time $s$

\begin{eqnarray}
\dot{G}_{k k l l}(s) &=& \exp(-(\Gamma_{\alpha \beta}+\Gamma_{\beta \alpha}) s)
(-\Gamma_{\alpha \beta} \delta_{k \alpha} \delta_{l \alpha}+\Gamma_{\beta \alpha} \delta_{k \alpha} \delta_{l \beta}
\nonumber
\\
&+&\Gamma_{\alpha \beta} \delta_{l \beta} \delta_{k \alpha}-\Gamma_{\beta \alpha} \delta_{k \beta} \delta_{l \beta}),
\end{eqnarray}

the population component of the response function $R_{2g}$ becomes

\begin{eqnarray}
\label{eq:R_2g_population}
R_{2g,pop}(t_1,t_2,t_3)&=&\sum_{\{k l\} \in \{\alpha,\beta\}}
e^{i \omega_{k g} t_1 - i \omega_{l g} t_3}
\\
&&e^{-\frac{1}{2}\Gamma_{k l} t_1 -\frac{1}{2}\Gamma_{l k} t_3}
\nonumber \\
&&e^{-g^{*}_{k k}(t_1)+g_{k l}(t_2)-g^{*}_{l l}(t_3)-g^{*}_{k l}(t_1+t_2)-g_{k l}(t_2+t_3)+g^{*}_{k l}(t_1+t_2+t_3)}
\nonumber \\
&&\int_{0}^{t_2} ds \dot{G}_{k k l l}(s) e^{2 i \Im(g_{l l}(t_2-s)-g_{k l}(t_2-s)+g_{k l}(t_2-s+t_3)-g_{l l}(t_2-s+t_3))}.
\nonumber
\end{eqnarray}

If all line shape functions are equal, independent of their indices, the argument of the exponential within the integrand vanishes.
Then the integral can be replaced by the relaxation tensor from Eq.~(\ref{eq:relaxation_dynamics}).

Pure dephasing between the electronic states is captured by the line shape function based formulation of the response function components. 
As in the cumulant expansion the electronic ground state is taken as the reference state, also pure dephasing between the singly excited states and the electronic ground state is contained, even though the latter does not appear as a line shape function index.
The dephasing rates due to lifetime broadening in response function components with coherence evolution between $\alpha$ and $\beta$ during $t_2$, which are given in the Appendix, enter in terms of the tensor elements \cite{PiMaFl06_234505_}

\begin{equation} \label{eq:dephasing_rates}
{\cal G}_{\alpha \beta \alpha \beta}(t_2)={\cal G}_{\beta \alpha \beta \alpha}(t_2)=\exp \left( -\frac{1}{2}(\Gamma_{\alpha \beta}+\Gamma_{\beta \alpha}) t_2 \right).
\end{equation}

Note that in the so-called modified Redfield approach for the intermediate regime between F\"{o}rster and Redfield limit \cite{YaFl_02_CP_163} the relaxation rates themselves depend on line shape functions and their derivatives.
By including this description, time-dependent relaxation rates can be obtained, as described in the Appendix.
However, this treatment only plays a role in cases where the line shape function parameters of the states involved in the relaxation process are different. 
Otherwise, the relaxation rates can be determined from the spectral density, which accounts for the influence of the environment.
In the modeling of systems like nanoparticles (quantum dots), where the relaxation process depends on both electronic structure and the phonon properties, phenomenological rate constants can be used for a simplified description \cite{SeHaPu13_JPCB}.

Approximate formulas for the limit of large population times relative to the time scale of vibrational relaxation of the bath components and under the assumption of uncorrelated fluctuations, i.e.\ $g_{i j}=0$ for $i \neq j$ \cite{AbMu11_JCP_174504}, are given in the literature \cite{Abramavicius09CR}. 
In this limit, one can assume that the asymptotic time derivative of the imaginary part of the line shape functions corresponds to the negative signed reorganization energy  

\begin{equation} \label{eq:reorganization_energy_limit}
-\lambda_{ll}=\Im{\lim_{t' \to \infty} \dot{g}_{ll}(t')}.
\end{equation}

The assumption of an asymptotically linear imaginary line shape function part $\Im(g_{ll}(t'))=-i \lambda_{l l} t'$ in the limit of a large time argument relative to the timescale of vibrational relaxation is supported by the justification of a linear approximation for the line shape function in the case of fast vibrational relaxation \cite{Mukamel95}. 

By assuming that vibrational relaxation has been completed, the line shape function part of the integrand from Eq.~(\ref{eq:R_2g_population}) becomes independent of the integration variable, so that

\begin{eqnarray}
\label{eq:R_2g_population_approximation}
R_{2g,pop}(t_1,t_2,t_3)&=&\sum_{\{k l\} \in \{\alpha,\beta\}}
e^{i \omega_{k g} t_1 - i \omega_{l g} t_3}
\nonumber \\
&&e^{-\frac{1}{2}\Gamma_{k l} t_1 -\frac{1}{2}\Gamma_{l k} t_3}G_{k k l l}(t_2)
\nonumber \\
&&e^{-g^{*}_{k k}(t_1)-g^{*}_{l l}(t_3)+2i \lambda_{l l} t_3}
\end{eqnarray}

is obtained.

Following an analogous scheme as in the derivation of Eq.~(\ref{eq:R_2g_population}), the population transfer components of the other SE and ESA response functions
can be derived, where the latter also include excitations involving the doubly excited state $f$.
The respective expressions are given in the Appendix together with contributions of the SE and ESA evolving in a coherence during the time interval $t_2$ and the ground state bleaching (GSB) response functions.
 
In all response functions the average electronic excitation energy of the singly excited states 
$\omega_{e g}=\frac{1}{2}(\omega_{\alpha g}+\omega_{\beta g})$ is subtracted from the oscillatory components with frequencies $\omega_{\alpha g}$ 
or $\omega_{\beta g}$ to allow an increase of the step size. This energetic shift determines the position of the origin of the two-dimensional spectrum. 

\subsection{Calculation of 2D-spectra}

In a conventional 2D-spectroscopy experiment, the coherence time $\tau$ between pulse $1$ and pulse $2$ is varied. The population time $T$ between the second incoming pulse (either pulse $1$ or pulse $2$) and pulse $3$ enters as a parameter. Furthermore, for a concise description the detection time $t$ between the interaction with the pulse $3$ and the signal detection is introduced, even though the corresponding frequency information stems from spectral resolution of the signal in experiment.
The indices of the pulses refer to the wavevector components in the chosen detection direction $\vec{k}_s=-\vec{k}_1+\vec{k}_2+\vec{k}_3$.
When finite pulse widths are not taken int account, $\tau$, $T$ and $t$ can be identified with $\pm t_1$, $t_2$ and $t_3$, where the positive or negative sign depends on whether a non-rephasing ($R_{1g}$, $R_{4g}$, $R^{*}_{2f}$) or a rephasing ($R_{2g}$, $R_{3g}$, $R^{*}_{1f}$) is considered.

For the calculation of a 2D-spectrum from the third-order polarization $P^{(3)}$ via two-dimensional Fourier transformation with respect to $t_1$ and $t_3$ the formula \cite{Kjellberg06PhysRevB}

\begin{eqnarray}
\label{eq:calculation_2D-spectrum}
\sigma_{2D,R}(\omega_{\tau},t_2,\omega_{t})&=&\int^{\infty}_{0} d t_1
\int^{\infty}_{0} d t_3
\nonumber
\\
&&\exp(\mp i \omega_{\tau} t_1) \exp(i \omega_{t} t_3) P^{(3)}(t_1,t_2,t_3)
\end{eqnarray}

is used, where the negative or positive sign in the complex exponential containing $t_1$ depends on whether $P^{(3)}$ consists of contributions from rephasing or nonrephasing response functions.

To obtain the response functions, the line shape function terms are calculated via \cite{Mukamel95}

\begin{eqnarray}
\label{eq:line_shape_function}
g_{i j}(t)&=&\frac{1}{2 \pi} \int^{\infty}_{-\infty} d \omega \frac{1-\cos(\omega t)}{\omega^2} \coth \left( \frac{\omega}{2 k_B T} \right) J_{i j}(\omega) 
\nonumber
\\
&+&\frac{i}{2 \pi} \int^{\infty}_{-\infty} d \omega \frac{\sin(\omega t)-\omega t}{\omega^2} J_{i j}(\omega), \; \{ i,j \} \in \{ \alpha,\beta,f \}
\end{eqnarray}

where the spectral density $J_{i j}(\omega)$ is composed of Debye and Lorentzian components in our model.

The Debye spectral density component, which is included to account for fluctuations of the enviroment, is given as \cite{SeHaPu13_JPCB,Kjellberg06PhysRevB}

\begin{equation}
\label{eq:Debye_spectral_density}
J_{D, i j}=2 \pi S_{D, i j} sgn(\omega) \frac{\omega^4}{2 \omega_c^3}\exp \left( -\frac{\omega}{\omega_c} \right), \; \{ i,j \} \in \{ \alpha,\beta,f \}
\end{equation}

while the Lorentzian spectral density component for the description of vibrations of the system reads \cite{BuVaAb12_JCP_044513}

\begin{equation}
\label{eq:Lorentzian_spectral_density}
J_{L, i j}=\frac{2 \sqrt{2} S_{L, i j} \omega^3_{0,L} \gamma_{L} \omega}{(\omega^2-\omega^2_{0,L})^2 + 2 \gamma^2_L \omega^2}, \; \{ i,j \} \in \{ \alpha,\beta,f \}.
\end{equation} 

The Huang-Rhys factors of the doubly excited state, which enter in $g_{f f}(t)$, $g_{f \alpha}(t)$ and $g_{f \beta}(t)$ do not generally need to be the same as for the singly excited state. This independence of the Huang-Rhys factors related to singly and doubly excited states mirrors the individual electron phonon coupling for single and double excitation. Note that the case of equal equilibrium distances between shifted harmonic oscillators involved in single and double excitation requires the Huang-Rhys factor of the doubly excited state to be four times larger than the one of the singly excited state.
This condition stems from the selection of the electronic ground state as the reference state in the cumulant expansion.

\section{Results}

\subsection{Choice of parameters and model assumptions}

Our model of two singly excited states with relaxation between them and one doubly excited state is very general and can be used to describe 2D spectra in many different systems like dye molecules or conjugated polymers with clearly laying electronic states. Excitonic dimers can be analyzed. Even molecules with pronounced vibronic progression can be treated if the ground state bleach component can be ignored. The two excited states would in this case correspond to two vibronic states. Here we have chosen a parameter set which shares most properties with the quantum dot model proposed in \cite{SeHaPu13_JPCB}.
The inhomogeneous broadening in terms of a particle size dependence of electronic excitation energies and vibrational frequencies is not taken into account. 
The energetic positions of the singly excited states relative to the electronic ground state are $\omega_{\alpha g}=\unit[17500]{cm^{-1}}$ and $\omega_{\beta g}=\unit[16500]{cm^{-1}}$. Different from the treatment in \cite{SeHaPu13_JPCB}, also the response function terms for the description of population transfer, which were derived in the previous section, are taken into account. The rate of relaxation from $\alpha$ to $\beta$ is chosen as $\Gamma_{\alpha \beta}=\unit[15]{cm^{-1}}$, which corresponds to a time constant of about $\unit[400]{fs}$. 
In the Debye spectral density component the parameters $S_{D}=\unit[0.25]{}$ and $\omega_{c}=\unit[25]{cm^{-1}}$ are taken as independent of the assigned singly excited state, whereas in the Lorentzian spectral density components the Huang-Rhys factors are assumed to depend on the singly excited electronic level.  
The assumption of different Huang-Rhys factors describes a more general situation than the assumption of equal Huang-Rhys factors, where in the case of correlated fluctuations, i.e.\ $g_{\alpha \beta}=g_{\alpha \alpha}=g_{\beta \beta}$, the integral expression in the response functions yields the relaxation tensor from
Eq.~(\ref{eq:relaxation_dynamics}). Therefore, different Huang-Rhys factors $S_{L,\alpha \alpha}=\unit[0.5]{}$ and $S_{L,\beta \beta}=\unit[1]{}$ are chosen. 
In the spectral density related to a coherence between different singly excited states, the Huang-Rhys factor is taken as 

\begin{equation} \label{eq:mixed_HR-factors}
S_{L,\alpha \beta}=S_{L,\beta \alpha}=\sqrt{S_{L,\alpha \alpha} S_{L,\beta \beta}}, 
\end{equation}

following the argumentation from \cite{NeMiMa10_094505_}. Under the assumption of uncorrelated fluctuations the respective Huang-Rhys factor is zero. The central frequency of $\omega_{0,L}=\unit[200]{cm^{-1}}$ and damping constant of $\gamma_{L}=\unit[25]{cm^{-1}}$ are assumed as independent of the singly excited level. Different vibrational frequencies could only be treated exactly by an approach beyond the second-order cumulant expansion \cite{FiEn13_JPCA_9444}.

To illustrate the significance of using the formulas for the response functions, which rigorously include the relaxation dynamics, the response function $R_2$ is considered in the following as an example.
Thereby only the population terms are taken into account to allow a direct comparison without regarding the influence of coherence terms. The latter are expected to decay on a much faster time scale than the one of population dynamics. The time scale of coherence dephasing is influenced by the amount of correlation between fluctuations in the coherently evolving electronic states \cite{CaZhDa14_JPCL_196}.
Also the difference between results from exact and approximative treatment of the population terms depends on assumptions about the correlations between fluctuations. In the following only the cases of perfectly correlated or uncorrelated fluctuations are considered for illustrative reasons.

\begin{figure}[h]
\centering
\includegraphics[width=10.0cm,height=6cm]{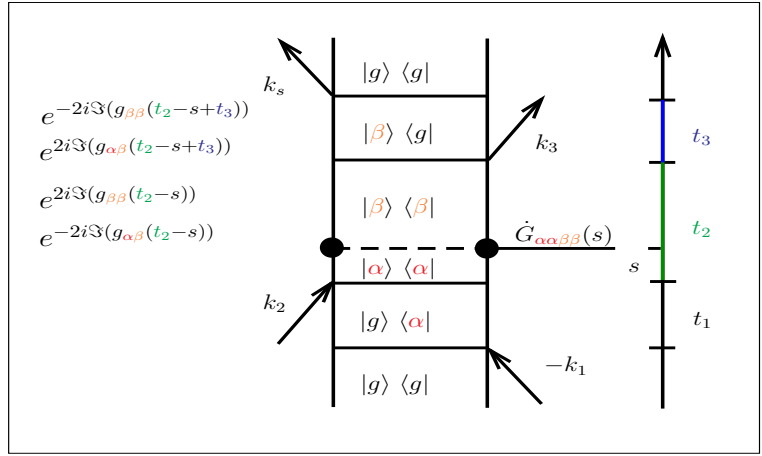}
\caption{\label{fig:illustration_Feynman_diagrams} Double sided Feynman diagram of the population transfer term of the response function $R_2$.}
\end{figure}

\begin{figure}[h]
\centering
\includegraphics[width=5.0cm,height=7.5cm]{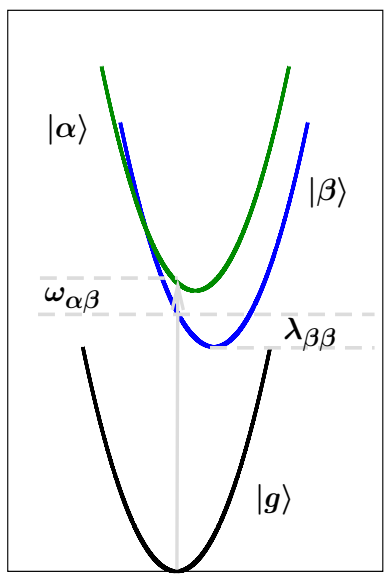}
\caption{\label{fig:sketch_potentials} Sketch of potential curves representing the Lorentzian oscillator mode to illustrate the combination of vibrational relaxation and relaxation between the singly excited sub-levels.}
\end{figure}

In Fig.~\ref{fig:illustration_Feynman_diagrams} the double sided Feynman diagrams of the population transfer term of response function $R_2$ is sketched. The effective relaxation rate at a selected value of the integration variable $s$ is both determined by the time derivative of the relaxation tensor $\dot{G}_{\alpha \alpha \beta \beta}(s)$ and the complex exponential factor containing positive and negative imaginary parts of the $s$-dependent
line shape functions. 
If the relaxation is considered as a stochastic process, $\dot{G}_{\alpha \alpha \beta \beta}(s) ds$ corresponds to the probability of the relaxation event to happen during $ds$.
If all line shape functions are equal and fluctuations are perfectly correlated, the latter factor becomes equal to one, so that only the first factor enters in the integration over $s$, yielding the relaxation tensor given in Eq.~(\ref{eq:relaxation_dynamics}).
If line shape functions with different indices vanish, i.e.\ under the assumption of uncorrelated fluctuations, Eq.~(\ref{eq:R_2g_population_approximation}) can be used to describe the limiting case of large enough population times, at which vibrational relaxation can be considered as complete.
Then the peak centers are expected to be shifted from their initial position at $t_2=\unit[0]{fs}$ to smaller energy in $\omega_t$-direction by an amount corresponding to twice the reorganization energy in the respective electronic state (see Fig.~\ref{fig:sketch_potentials}).
For a Lorentzian spectral density, the reorganization energy $\lambda_{ll}$ corresponds to the product of the central frequency $\omega_{0,L}$ and the Huang-Rhys factor $S_{L,l l}$ in the singly excited state $l \in \{ \alpha,\beta \}$.

\subsection{Comparison of results from exact and approximative calculations}

An overview of the influence of different model assumptions on the resulting 2D spectra is given in Table \ref{tab:overview_different_cases}. The results depend on whether fluctuations in the two singly excited states are taken as correlated or uncorrelated. Furthermore, the line shape function parameters play a role. 
For completeness, the specific case of equal Huang-Rhys factors in the Lorentzian spectral density components of both singly excited states under the assumption of correlated fluctuations, where the integral in Eq.~(\ref{eq:R_2g_population}) yields the relaxation tensor, enters in the Table \ref{tab:overview_different_cases}.
However, as the chosen parameters reflect the more general case of different Huang-Rhys factors, this specific case does not play a role in the discussion of the results from the model calculations, where the results from Eq.~(\ref{eq:R_2g_population}) are compared to those from an approximate treatment.
The approximation either consists in replacing the integral in Eq.~(\ref{eq:R_2g_population}) by the relaxation tensor from Eq.~(\ref{eq:relaxation_dynamics}) or in using the long time limit Eq.~(\ref{eq:R_2g_population_approximation}) instead of Eq.~(\ref{eq:R_2g_population}). The figures which show the respective comparisons are indicated in the rightmost column of the table.

\begin{figure}[h]
\centering
\includegraphics[width=10.0cm,height=12cm]{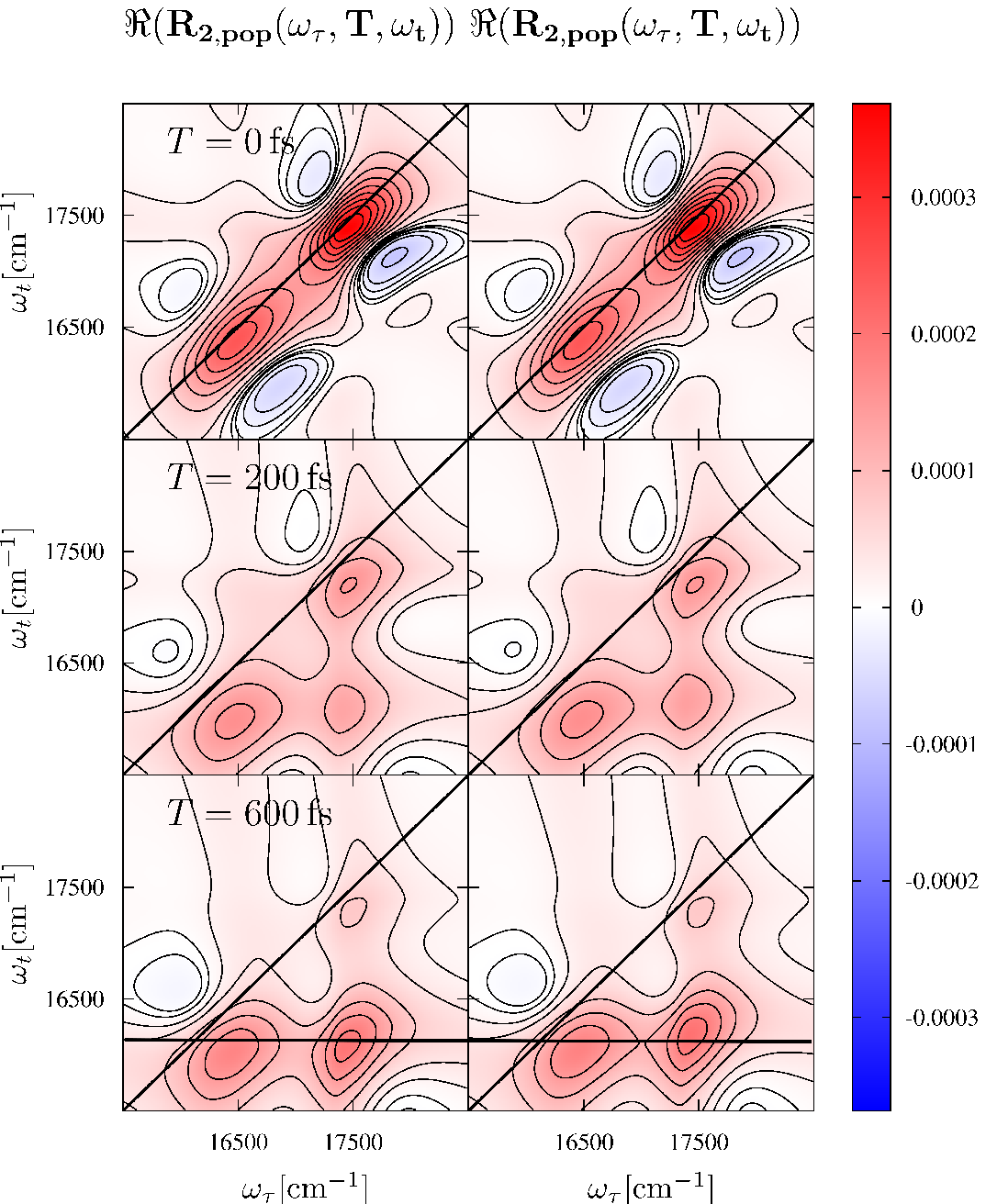}
\caption{\label{fig:plot_R2_real_correlated_fluctuations_with_and_without_integration} Real part of the 2D-spectra of $R_2$ at different population times in the case of correlated fluctuations with explicit integration in Eq.~(\ref{eq:R_2g_population}) (left hand side) and with replacement of the integral by the relaxation tensor from Eq.~(\ref{eq:relaxation_dynamics}) (right hand side).
In the lowest row, the horizontal line indicates the position $\omega_t=\omega_{\beta g}-2\lambda_{\beta \beta}$.}
\end{figure}

For the calculation of the results shown in Fig.~\ref{fig:plot_R2_real_correlated_fluctuations_with_and_without_integration} correlated fluctuations were assumed.
The 2D-spectra for population times $T=\unit[0]{fs}$, $T=\unit[200]{fs}$ and $T=\unit[600]{fs}$ in the left column were calculated from Eq.~(\ref{eq:R_2g_population}), whereas the corresponding 2D-spectra in the right column stem from calculations with replacement of the integral by the relaxation tensor from Eq.~(\ref{eq:relaxation_dynamics}). 
As the chosen value of $\gamma_{L}$ leads to relatively fast vibrational relaxation with a time constant of about $\unit[200]{fs}$, the maximum of the population transfer crosspeak appears at $\omega_t$ positions below the vertical transition energy of the lower singly excited state $\omega_{\beta g}$ already at $T=\unit[200]{fs}$. With further increasing population time the crosspeak maximum is shifted further towards its final $\omega_t$ position close to $\omega_{\beta g}-2\lambda_{\beta \beta}$, which can be predicted by assuming the line shape functions as linear in the limit of population time to infinity.
More precisely, the line shape function components from Eq.~(\ref{eq:R_2g_population}) outside the integral with dependence on $t_2$ lead to a factor 
$\exp(2 i \lambda_{\alpha \beta} t_3)$ in the limit of linear Lorentzian components. In this limit of large $t_2$ relative to the timescale of vibrational relaxation, the line shape functions in the exponential factor of the integrand becomes independent of the integration variable, yielding a factor $\exp(2 i (\lambda_{\beta \beta}-\lambda_{\alpha \beta}) t_3)$. This factor describes the $t_3$-dependence of the integral expression under the assumption that oscillations from the complex exponential line shape function expression result in cancellation of the integral until vibrational relaxation has completely taken place. 
Note that not only the Lorentzian but also the Debye component of the line shape functions yields a reorganization energy contribution, so that the resulting peak position slightly differs from $\omega_{\beta g}-2\lambda_{\beta \beta}$ if only the reorganization energy of the Lorentzian spectral density component is considered. 
With increasing value of $t_2$ the relative crosspeak intensity increases, which indicates population transfer from $\alpha$ to $\beta$. 
The differences between the results from exact and approximate calculation are mainly related to the maximum position of the population transfer crosspeak. In the latter case, where the combined dynamics of vibrational relaxation and population transfer are not treated properly, the $\omega_t$ position of the respective peak maximum tends towards $\omega_t=\omega_{\beta g}-2\lambda_{\alpha \beta}$, as the integral expression yields no frequency shift.
Therefore, the influence of vibrational relaxation is underestimated in the approximate treatment.

\begin{figure}[h]
\centering
\includegraphics[width=10.0cm,height=12cm]{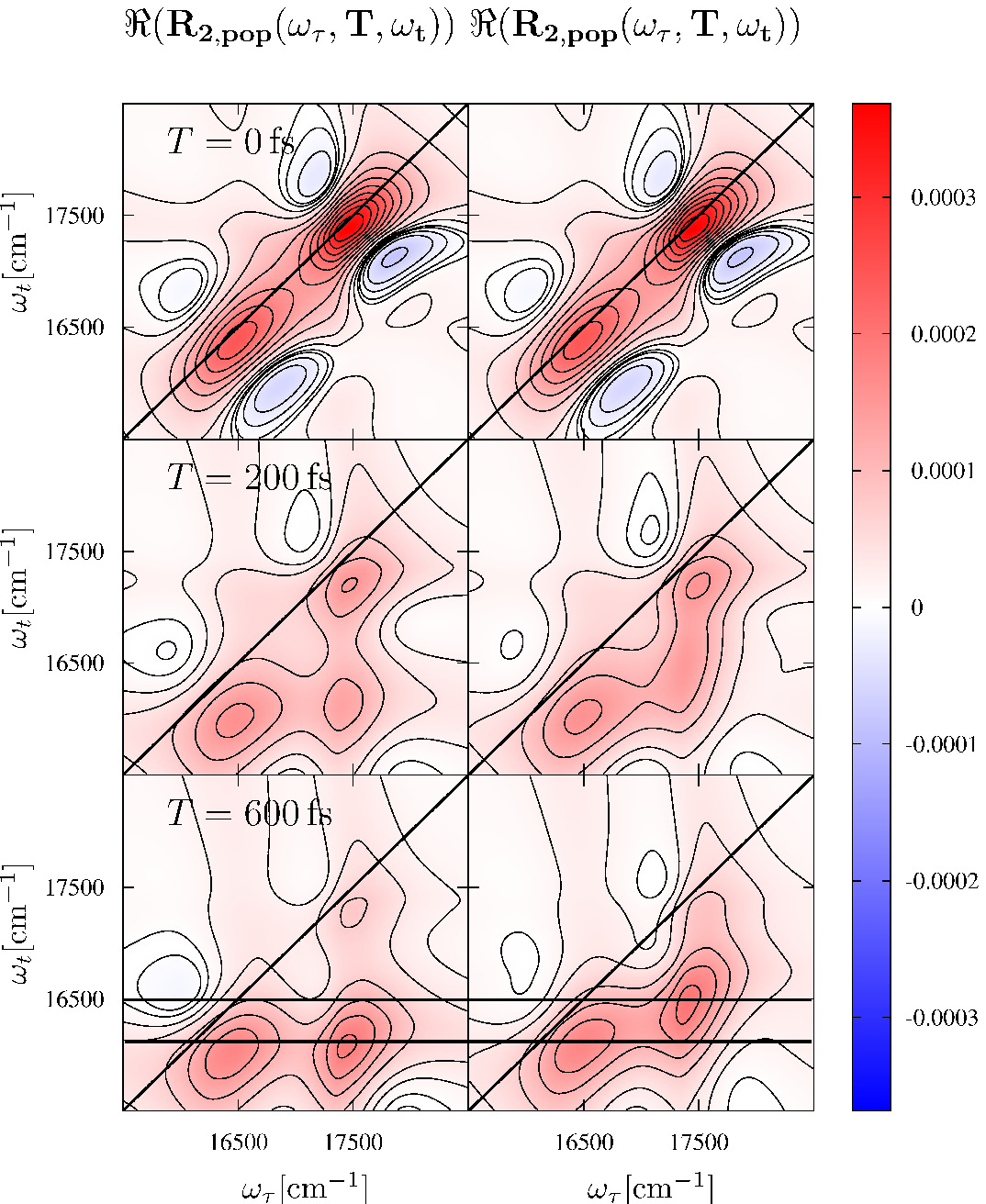}
\caption{\label{fig:plot_R2_real_uncorrelated_fluctuations_with_and_without_integration} Real part of the 2D-spectra of $R_2$ at different population times in the case of uncorrelated fluctuations with explicit integration in Eq.~(\ref{eq:R_2g_population}) (left hand side) and with replacement of the integral by the relaxation tensor from Eq.~(\ref{eq:relaxation_dynamics}) (right hand side).
In the lowest row, the horizontal line indicates the positions $\omega_t=\omega_{\beta g}$ and $\omega_t=\omega_{\beta g}-2\lambda_{\beta \beta}$.}
\end{figure}

This effect is even more pronounced in the case of uncorrelated fluctuations shown in Fig.~\ref{fig:plot_R2_real_uncorrelated_fluctuations_with_and_without_integration}, where the line shape functions with mixed indices are taken as zero. 
This assumption is equivalent to a reorganization energy $\lambda_{\alpha \beta}$ of zero, so that without rigorous integration over the combined dynamics of fluctuations and population transfer, the population transfer crosspeak ends up at the vertical transition energy $\omega_t=\omega_{\beta g}$, independent of an increase of $t_2$.
This effect indicates that vibrational relaxation is not taken into account at all in the description of the population transfer term when the integral expression in Eq.~(\ref{eq:R_2g_population}) is approximated by the relaxation tensor. Thus, in the case of uncorrelated fluctuation, this approximation is not appropriate.

\begin{figure}[h]
\centering
\includegraphics[width=10.0cm,height=8cm]{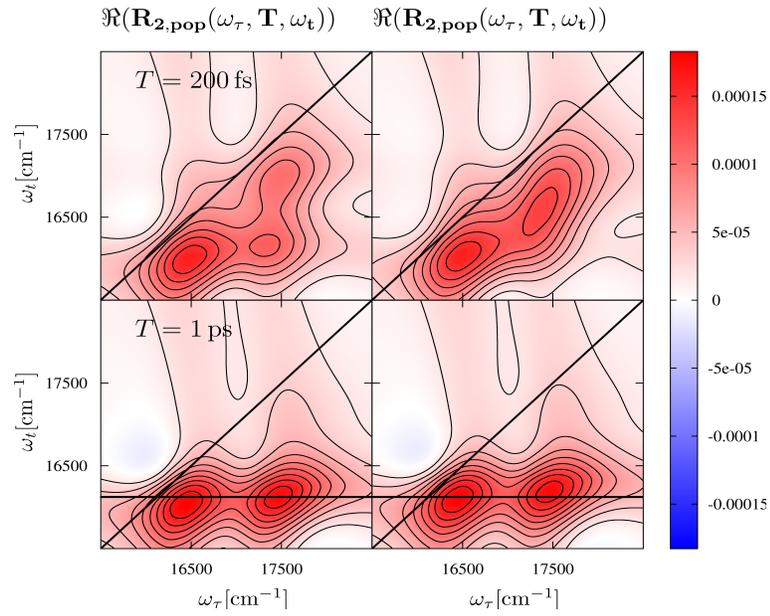}
\caption{\label{fig:plot_R2_real_uncorrelated_fluctuations_limiting_case_with_and_without_integration} Real part of the 2D-spectra of $R_2$ at $T=\unit[200]{fs}$ and $T=\unit[1]{ps}$, calculated according to Eq.~(\ref{eq:R_2g_population}) (left hand side) and using the approximation from Eq.~(\ref{eq:R_2g_population_approximation}) (right hand side).
In the lowest row, the horizontal line indicates the position $\omega_t=\omega_{\beta g}-2\lambda_{\beta \beta}$.}
\end{figure}

However, under the assumption of uncorrelated fluctuations and in the limit of large population times relative to the timescalce of dissipative dynamics, the results from Eq.~(\ref{eq:R_2g_population}) are expected to resemble the ones obtained by using Eq.~(\ref{eq:R_2g_population_approximation}), which describes the situation where vibrational relaxation has been completed. 
If the latter assumption is not valid yet, the approximation leads to incorrect shapes and positions of all appearing peaks, as shown for $T=\unit[200]{fs}$ 
in the first row of Fig.~\ref{fig:plot_R2_real_uncorrelated_fluctuations_limiting_case_with_and_without_integration}. The 2D-spectrum on the left hand side stems from a calculation using Eq.~(\ref{eq:R_2g_population}), the one at the right hand side was obtained from Eq.~(\ref{eq:R_2g_population_approximation}).
In the limit of population times where vibrational relaxation has been completed, both ways of calculating the 2D-spectra yield similar results, as displayed for $T=\unit[1]{ps}$ in the lower row. For the remaining differences in the peak positions with respect to $\omega_t$, the influence of the Debye spectral density component plays a role.

\subsection{Comparison of peak evolution under the assumptions of correlated and uncorrelated fluctuations}

\begin{figure}[h]
\centering
\includegraphics[width=10.0cm,height=6cm]{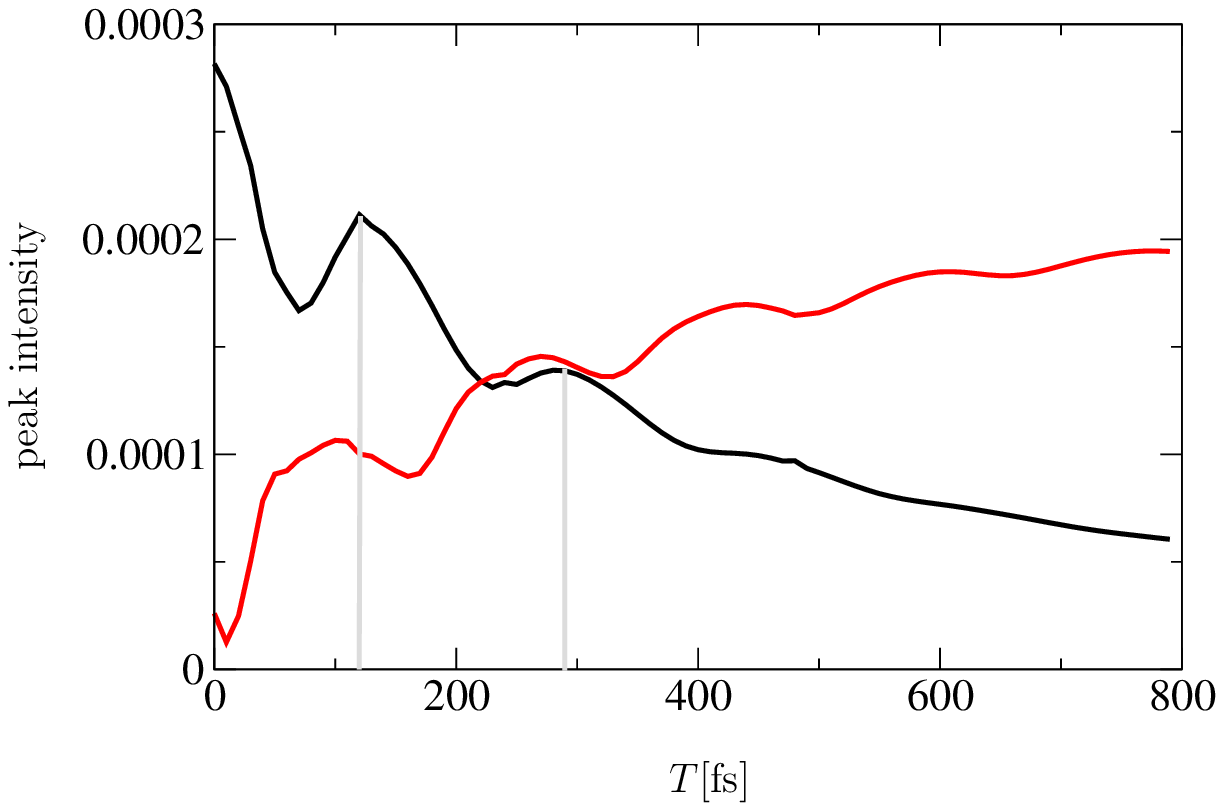}
\caption{\label{fig:time_evolution_diagonal_peak_crosspeak_R2_real_correlated_fluctuations} Population time evolution of the real part of the 2D-spectrum of $R_2$  at ($\omega_{\tau}=\unit[17450]{cm^{-1}}$,$\omega_{\tau}=\unit[17240]{cm^{-1}}$) (black curve) 
and at ($\omega_{\tau}=\unit[17450]{cm^{-1}}$,$\omega_{\tau}=\unit[15970]{cm^{-1}}$) (red curve) in the case of correlated fluctuations.}
\end{figure}

\begin{figure}[h]
\centering
\includegraphics[width=10.0cm,height=6cm]{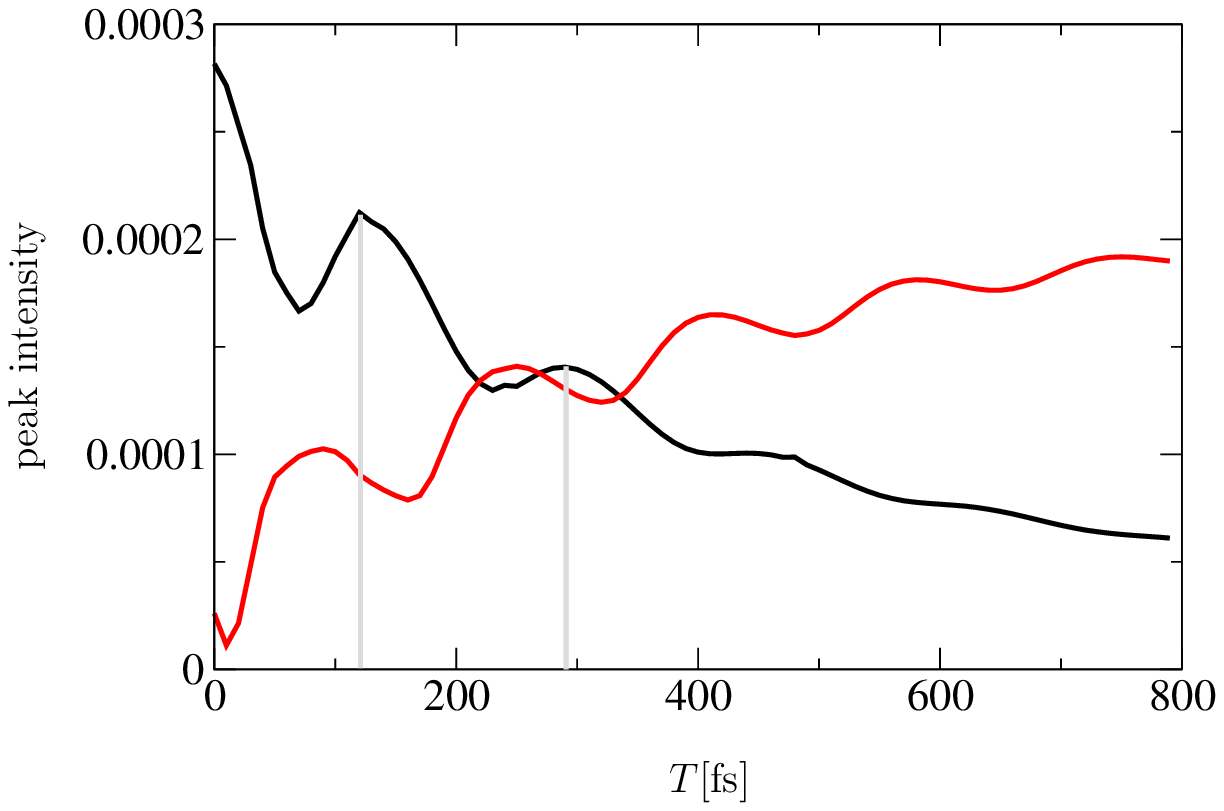}
\caption{\label{fig:time_evolution_diagonal_peak_crosspeak_R2_real_uncorrelated_fluctuations} Population time evolution of the real part of the 2D-spectrum of $R_2$  at ($\omega_{\tau}=\unit[17450]{cm^{-1}}$,$\omega_t=\unit[17240]{cm^{-1}}$) (black curve) 
and at ($\omega_{\tau}=\unit[17450]{cm^{-1}}$,$\omega_t=\unit[15970]{cm^{-1}}$) (red curve) in the case of uncorrelated fluctuations.}
\end{figure}

In Fig.~\ref{fig:time_evolution_diagonal_peak_crosspeak_R2_real_correlated_fluctuations} and Fig.~\ref{fig:time_evolution_diagonal_peak_crosspeak_R2_real_uncorrelated_fluctuations} the evolution of the real part of the 2D-spectrum at selected points as a function of the population time is shown for the case of correlated fluctuations and uncorrelated fluctuations, respectively.
For the calculation of the 2D-spectra, Eq.~(\ref{eq:R_2g_population}) was used in both cases.
The black curve is related to the evolution at position ($\omega_{\tau}=\unit[17450]{cm^{-1}}$,$\omega_{\tau}=\unit[17240]{cm^{-1}}$) in the region of the upper diagonal peak, the red curve belongs to the point ($\omega_{\tau}=\unit[17450]{cm^{-1}}$,$\omega_{\tau}=\unit[15970]{cm^{-1}}$) in the crosspeak region.
While the evolution of the black curve is almost the same in both cases, a phase shift of the vibrational oscillations in the red curve appears, depending on whether correlated or uncorrelated fluctuations are assumed. For correlated fluctuations the positions of the selected local maxima of the black curve almost coincide with the positions where local maxima of the red curve appear. In contrast, for uncorrelated fluctuations the red curve exhibits inflection points at the respective positions.
More precisely, as long as vibrational relaxation cannot be considered as completed, the complex exponential containing line shape function in the integrand from Eq.~(\ref{eq:R_2g_population}) leads to oscillations with dependence on the integation variable. Therefore, the integral yields an oscillatory component with respect to $t_2$, which shifts the frequency of peak oscillations with respect to $T$ at a selected ($\omega_{\tau}$,$\omega_t$) position. 
Note that it is not possible to identify a phase shift independent of the selected position, as the oscillatory component of the integral expression depends on $t_3$ and, thus, on $\omega_t$.
However, the finding of a possible phase shift of about $\pi/2$ between diagonal- and crosspeaks in our calculation indicates that the interpretation of such a phase shift in terms of quantum transport \cite{PaVoAb11_PNAS_20908}, i.e.\ non-secular effects with conversion between electronic coherences and populations, cannot be generalized. In the context of our model assumptions the explanation of this phase shift rather amounts to mutual influence between vibrational coherence dynamics and the population transfer process between the singly excited electronic states. 

\section{Conclusions}

By combining the concepts of line shape functions and rate equations, we have derived third-order response functions of a model system including a term for population transfer between two singly excited states subsequent to excitation from the electronic ground state. 
The line shape function components are related to a composed spectral density including a Lorentzian component to account for vibrational effects, the relaxation rates are taken as constant.
Under the assumption of equal line shape function parameters and correlated fluctations in the singly excited levels, the response functions can be factorized in terms of line shape function components and relaxation tensor elements.
Different line shape function parameters, uncorrelated fluctuations or a combination of both lead to a non-trivial integral expression, which accounts for the interplay between fluctuation and relaxation dynamics. 
For both correlated and uncorrelated fluctuations, differences in the population transfer crosspeaks of 2D-spectra, calculated either from our derived formula or in an approximative way, were discussed. We point out that the combined treatment of population transfer and fluctuation dynamics is required to properly account for the influence of the reorganization energy. The shortcoming of the approximation of decoupled population transfer and fluctuation dynamics is less pronounced for correlated than for uncorrelated fluctuations.
However, in the latter case a different approximation can be used to obtain agreement with the exact calculation in the limit of large population times relative to the fluctuation decay time scale. By considering the vibrational relaxation as completed, the line shape function component of the mixed dynamics becomes constant, so that integration is only related to the relaxation dynamics.
As a result, we conclude that our derived approach is useful in the intermediate regime, where relaxation and fluctuation dynamics cannot be separated from each other.
In this regime the time evolution of selected points in diagonal- and crosspeak region can lead to a different shift of the relative phase of vibrational oscillations, depending on the assumption of correlated and uncorrelated fluctuations. The phase shift reflects the mutual influence of population transfer and vibrational dynamics.
Our derived approach can be combined with the modified Redfield description, in principle.
Then the relaxation rates become time-dependent, so that the assumptions about correlations are expected to influence the population transfer crosspeak evolution to an even larger extent than in the presented study. 

The combination of our approach with the modified Redfield description would take all orders of the system-bath interaction into account, however in the framework of a Markovian description. Therefore such a combined treatment yields a benchmark for approximate methods, such as HEOM, in the case of a Markovian bath. If non-Markovian effects play a role, appearing differences in the population time evolution of the calculated 2D-spectra can be related to these effects. In this way it would be possible to identify features from non-Markovian effects in 2D-spectra.

In combination with the modified Redfield description, our derived approach could be particularly useful to calculate 2D spectra of a system with dynamics through a conical intersection subsequent to electronic excitation \cite{SeEi12_JCP_024109,TiPeJo13_PNAS_1203,KiFeTi14_JCP}. If a line shape function description is chosen for the dynamics at the conical intersection \cite{SmFaJo05_JCP}, population and coherence terms of the response functions can be related to the evolution along the tuning and coupling coordinate, respectively \cite{PeSmJo11}.

\section*{Acknowledgement}
We gratefully acknowledge financial support of the Knut and Alice Wallenberg Foundation, Swedish Research Council, and Swedish Energy Agency.
Collaboration within nmC@LU is acknowledged.

\begin{appendix}

\section{Combination of the derived approach with the modified Redfield description}

If the population transfer kinetics between the singly excited states with indices $\{k,l\} \in \{\alpha,\beta\}$ is assumed to exclusively depend on the bath dynamics, which is described by line shape functions, the respective rate constants $R_{k k l l}$, which are equivalent to $\Gamma_{k l}$, can be calculated by using the formula \cite{ZhMeCh98_JCP_7763} 

\begin{eqnarray}
\label{eq:modified_Redfield rate}
\Gamma_{k l}&=&R_{k k l l}=-2 \Re \int_{0}^{\infty} dt' \exp(-i (\omega_{k g}-\omega_{l g}) t' - g_{k k, k k}(t') - g_{l l, l l}(t') 
\nonumber
\\
&+& g_{l l, k k}(t') + g_{k k, l l}(t') - 2i (\lambda_{l l, l l}-\lambda_{k k, l l}) t') 
\nonumber
\\
&& \times \{ \ddot{g}_{k l, l k}(t') - (\dot{g}_{l k, l l}(t')-\dot{g}_{l k, k k}(t')+2i \lambda_{l k, l l})
\nonumber
\\
&&\times (\dot{g}_{l l, l k}(t')-\dot{g}_{k k, k l}(t')+2i \lambda_{k l, l l}) \}.
\end{eqnarray}

As in our model the indices $\{k,l\} \in \{\alpha,\beta\}$ are related to eigenstates, the two index pairs of the line shape functions are related to correlations between the fluctuations in the respective eigenstates \cite{NoPaAm04_JCPB_10363}. These correlations scale with the Huang-Rhys factors of the included spectral density components. Therefore, the Lorentzian line shape function components with two index pairs from Eq.~(\ref{eq:modified_Redfield rate}) can be taken as

\begin{equation}
g_{k l, m n}(t')=\sqrt{\frac{S_{L,k l} S_{L,m n}}{S_{L,\alpha \alpha} S_{L,\alpha \alpha}}} g_{\alpha \alpha}(t'), \; \{k,l,m,n\} \in \{\alpha,\beta\}
\end{equation}

in our model, provided that $S_{L,\alpha \alpha}$ is larger than zero. Huang-Rhys factors with different indices are calculated according to Eq.~(\ref{eq:mixed_HR-factors}) in the case of correlated fluctuations, while they are zero under the assumption of uncorrelated fluctuations.
For a Lorentzian oscillator the reorganization can be calculated as 

\begin{equation}
\lambda_{k l, m n}(t')=\sqrt{\frac{S_{L,k l} S_{L,m n}}{S_{L,\alpha \alpha} S_{L,\alpha \alpha}}} \lambda_{\alpha \alpha}(t'), \; \{k,l,m,n\} \in \{\alpha,\beta\},
\end{equation}

otherwise it can be obtained from the asymptotic time derivative of the respective line shape function according to Eq.~(\ref{eq:reorganization_energy_limit}). 

In the calculation of the population transfer rate from Eq.~(\ref{eq:modified_Redfield rate}) the integration leads to cancellation of oscillatory components, so that only the dissipative dynamics remains. To obtain time-dependent rates, the limit of the upper integration border to infinity can be replaced by the respective time variable.
For time-dependent relaxation rates, the relaxation tensor can be obtained by solving the differential equation

\begin{equation} \label{eq:integration_relaxation_tensor}
\dot{G}_{k k l l} \left( s \right) = -\sum_{m \in \{\alpha,\beta\}} R_{k k m m}\left( s \right) G_{m m l l}(s),
\end{equation}

where $R_{m m l l}(s)$ is determined from Eq.~(\ref{eq:modified_Redfield rate}) with upper integration border $s$ instead of $\infty$.
The initial condition for the solution of Eq.~(\ref{eq:integration_relaxation_tensor}) is $G_{k k m m}(0)=\delta_{k m}, \; \{k,m\} \in \{\alpha,\beta\}$.
By inserting the time-dependent relaxation rates into Eq.~(\ref{eq:dephasing_rates}), the dephasing tensor elements can be obtained.

\section{Response functions}

Following the notation for the different response functions from \cite{SeHaPu13_JPCB}, the population components of the SE and ESA response functions, which stem from an analogous derivation as Eq.~(\ref{eq:R_2g_population}), are obtained as

\begin{eqnarray}
\label{eq:R_1g_population_omega_e_g_subtracted}
R_{1g,pop}(t_1,t_2,t_3)&=&\sum_{\{k l\} \in \{\alpha,\beta\}}
e^{- i (\omega_{k g}-\omega_{e g}) t_1 - i (\omega_{l g}-\omega_{e g}) t_3}
\\
&&e^{-\frac{1}{2}\Gamma_{k l} t_1 -\frac{1}{2}\Gamma_{l k} t_3}
\nonumber \\
&&e^{-g_{k k}(t_1)-g^{*}_{k l}(t_2)-g^{*}_{l l}(t_3)+g_{k l}(t_1+t_2)+g^{*}_{k l}(t_2+t_3)-g_{k l}(t_1+t_2+t_3)}
\nonumber \\
&&\int_{0}^{t_2} ds \dot{G}_{k k l l}(s) e^{2 i \Im(g_{l l}(t_2-s)-g_{k l}(t_2-s)+g_{k l}(t_2-s+t_3)-g_{l l}(t_2-s+t_3))},
\nonumber 
\end{eqnarray}

\begin{eqnarray}
\label{eq:R_2g_population_omega_e_g_subtracted}
R_{2g,pop}(t_1,t_2,t_3)&=&\sum_{\{k l\} \in \{\alpha,\beta\}}
e^{i (\omega_{k g}-\omega_{e g}) t_1 - i (\omega_{l g}-\omega_{e g}) t_3}
\\
&&e^{-\frac{1}{2}\Gamma_{k l} t_1 -\frac{1}{2}\Gamma_{l k} t_3}
\nonumber \\
&&e^{-g^{*}_{k k}(t_1)+g_{k l}(t_2)-g^{*}_{l l}(t_3)-g^{*}_{k l}(t_1+t_2)-g_{k l}(t_2+t_3)+g^{*}_{k l}(t_1+t_2+t_3)}
\nonumber \\
&&\int_{0}^{t_2} ds \dot{G}_{k k l l}(s) e^{2 i \Im(g_{l l}(t_2-s)-g_{k l}(t_2-s)+g_{k l}(t_2-s+t_3)-g_{l l}(t_2-s+t_3))},
\nonumber
\end{eqnarray}

\begin{eqnarray}
\label{eq:R_1fcc_population_omega_e_g_subtracted}
R^{*}_{1f,pop}(t_1,t_2,t_3)&=&\sum_{\{k l\} \in \{\alpha,\beta\}}
e^{i (\omega_{k g}-\omega_{e g}) t_1 - i (\omega_{f l}-\omega_{e g}) t_3}
\\
&&e^{-\frac{1}{2}\Gamma_{k l} t_1 -\frac{1}{2}\Gamma_{l k} t_3}
\nonumber \\
&&e^{-g^{*}_{k k}(t_1)-g_{k l}(t_2)-g_{l l}(t_3)+g^{*}_{k l}(t_1+t_2)+g_{k l}(t_2+t_3)-g^{*}_{k l}(t_1+t_2+t_3)}
\nonumber \\
&&e^{+g_{f k}(t_2)+2g_{f l}(t_3)-g^{*}_{f k}(t_1+t_2)-g_{f k}(t_2+t_3)+g^{*}_{f k}(t_1+t_2+t_3)-g_{f f}(t_3)}
\nonumber \\
&&\int_{0}^{t_2} ds \dot{G}_{k k l l}(s) e^{2 i \Im(g_{k l}(t_2-s)-g_{l l}(t_2-s)+g_{l l}(t_2-s+t_3)-g_{k l}(t_2-s+t_3))}
\nonumber \\
&&e^{2 i \Im(g_{f l}(t_2-s)-g_{f k}(t_2-s)+g_{f k}(t_2-s+t_3)-g_{f l}(t_2-s+t_3))}
\nonumber 
\end{eqnarray}

and

\begin{eqnarray}
\label{eq:R_2fcc_population_omega_e_g_subtracted}
R^{*}_{2f,pop}(t_1,t_2,t_3)&=&\sum_{\{k l\} \in \{\alpha,\beta\}}
e^{- i (\omega_{k g}-\omega_{e g}) t_1 - i (\omega_{f l}-\omega_{e g}) t_3}
\\
&&e^{-\frac{1}{2}\Gamma_{k l} t_1 -\frac{1}{2}\Gamma_{l k} t_3}
\nonumber \\
&&e^{-g_{k k}(t_1)+g^{*}_{k l}(t_2)-g_{l l}(t_3)-g_{k l}(t_1+t_2)-g^{*}_{k l}(t_2+t_3)+g_{k l}(t_1+t_2+t_3)}
\nonumber \\
&&e^{-g^{*}_{f k}(t_2)+2g_{f l}(t_3)+g_{f k}(t_1+t_2)+g^{*}_{f k}(t_2+t_3)-g_{f k}(t_1+t_2+t_3)-g_{f f}(t_3)}
\nonumber \\
&&\int_{0}^{t_2} ds \dot{G}_{k k l l}(s) e^{2 i \Im(g_{k l}(t_2-s)-g_{l l}(t_2-s)+g_{l l}(t_2-s+t_3)-g_{k l}(t_2-s+t_3))}
\nonumber \\
&&e^{2 i \Im(g_{f l}(t_2-s)-g_{f k}(t_2-s)+g_{f k}(t_2-s+t_3)-g_{f l}(t_2-s+t_3))}.
\nonumber 
\end{eqnarray}

The coherence components of the SE and ESA response functions read

\begin{eqnarray} 
\label{eq:R_1g_coherence}
R_{1g,coh}(t_1,t_2,t_3)&=&\sum_{\{k l\} \in \{\alpha,\beta\}, k\neq l}
e^{- i (\omega_{k g}-\omega_{e g}) t_1 - i (\omega_{k g}-\omega_{l g}) t_2 - i (\omega_{k g}-\omega_{e g}) t_3}
\\
&&e^{-\frac{1}{2}\Gamma_{k l} t_1 - \frac{1}{2}\Gamma_{k l} t_3} G_{k l k l}(t_2)
\nonumber \\
&&e^{-g_{k l}(t_1)-g_{l l}^{*}(t_2)-g_{k l}^{*}(t_3)+g_{k l}(t_1+t_2)+g_{k l}^{*}(t_2+t_3)-g_{k k}(t_1+t_2+t_3)} 
\nonumber \\
\label{eq:R_2g_coherence}
R_{2g,coh}(t_1,t_2,t_3)&=&\sum_{\{k l\} \in \{\alpha,\beta\}, k\neq l}
e^{i (\omega_{k g}-\omega_{e g}) t_1 - i (\omega_{l g}-\omega_{k g}) t_2 - i (\omega_{l g}-\omega_{e g}) t_3}
\\
&&e^{-\frac{1}{2}\Gamma_{k l} t_1 -\frac{1}{2}\Gamma_{l k} t_3} G_{k l k l}(t_2)
\nonumber \\
&&e^{-g_{k l}^{*}(t_1)+g_{k l}(t_2)-g_{k l}^{*}(t_3)-g_{k k}^{*}(t_1+t_2)-g_{l l}(t_2+t_3)+g_{k l}^{*}(t_1+t_2+t_3)}  
\nonumber \\
\label{eq:R_1fcc_coherence}
R^{*}_{1f,coh}(t_1,t_2,t_3)&=&\sum_{\{k l\} \in \{\alpha,\beta\}, k\neq l}
e^{i (\omega_{k g}-\omega_{e g}) t_1 + i (\omega_{k g}-\omega_{l g}) t_2 
- i (\omega_{f k}-\omega_{e g}) t_3}
\\
&&e^{-\frac{1}{2}\Gamma_{k l} t_1-\frac{1}{2}\Gamma_{k l} t_3}G_{k l k l}(t_2)
\nonumber \\
&&e^{-g^{*}_{k l}(t_1)-g_{l l}(t_2)-g_{k l}(t_3)+g^{*}_{k l}(t_1+t_2)+g_{k l}(t_2+t_3)-g^{*}_{k k}(t_1+t_2+t_3)}
\nonumber \\
&&e^{+g_{l f}(t_2)+g_{l f}(t_3)+g_{k f}(t_3)-g^{*}_{k f}(t_1+t_2)-g_{l f}(t_2+t_3)+g^{*}_{k f}(t_1+t_2+t_3)-g_{f f}(t_3)}
\nonumber \\
\label{eq:R_2fcc_coherence}
R^{*}_{2f,coh}(t_1,t_2,t_3)&=&\sum_{\{k l\} \in \{\alpha,\beta\}, k\neq l}
e^{- i (\omega_{k g}-\omega_{e g}) t_1 - i (\omega_{k g}-\omega_{l g}) t_2 
- i (\omega_{f l}-\omega_{e g}) t_3}
\\
&&e^{-\frac{1}{2}\Gamma_{k l} t_1-\frac{1}{2}\Gamma_{l k} t_3}G_{k l k l}(t_2)
\nonumber \\
&&e^{-g_{k l}(t_1)+g^{*}_{k l}(t_2)-g_{k l}(t_3)-g_{k k}(t_1+t_2)-g^{*}_{l l}(t_2+t_3)+g_{k l}(t_1+t_2+t_3)}
\nonumber \\
&&e^{-g^{*}_{l f}(t_2)+g_{l f}(t_3)+g_{k f}(t_3)+g_{k f}(t_1+t_2)+g^{*}_{l f}(t_2+t_3)-g_{k f}(t_1+t_2+t_3)-g_{f f}(t_3)}
\nonumber 
\end{eqnarray}

The GSB contributions are given as

\begin{eqnarray} 
\label{eq:R_3g_coherence}
R_{3g}(t_1,t_2,t_3)&=&\sum_{\{k l\} \in \{\alpha,\beta\}}
e^{i (\omega_{k g}-\omega_{e g}) t_1 - i (\omega_{l g}-\omega_{e g}) t_3}
e^{- \frac{1}{2} \Gamma_{k l} t_1 - \frac{1}{2} \Gamma_{l k} t_3} 
\\
&&e^{-g_{k k}^{*}(t_1)+g_{k l}^{*}(t_2)-g_{l l}(t_3)-g_{k l}^{*}(t_1+t_2)-g_{k l}^{*}(t_2+t_3)+g_{k l}^{*}(t_1+t_2+t_3)}  
\nonumber \\
\label{eq:R_4g_coherence}
R_{4g}(t_1,t_2,t_3)&=&\sum_{\{k l\} \in \{\alpha,\beta\}}
e^{- i (\omega_{k g}-\omega_{e g}) t_1 - i (\omega_{l g}-\omega_{e g}) t_3}
e^{- \frac{1}{2} \Gamma_{k l} t_1 - \frac{1}{2} \Gamma_{l k} t_3} 
\\
&&e^{-g_{k k}(t_1)-g_{k l}(t_2)-g_{l l}(t_3)+g_{k l}(t_1+t_2)+g_{k l}(t_2+t_3)-g_{k l}(t_1+t_2+t_3)}.
\nonumber 
\end{eqnarray}

\end{appendix}

\newpage

\begin{table}[h]
\begin{center}
\begin{tabular}{| p{2.25cm} | p{3.75cm} | p{3cm} | p{3cm}| }
  \hline
  assumption about fluctuations & assumption about line shape functions & 
  kind of approximation & exact calculation vs.~approximation \\
  \hline
  \hline
  \multirow{2}{*}{correlated} & \vfill $g_{\alpha \beta}=g_{\alpha \alpha}=g_{\beta \beta}$ \vfill & 
  \multirow{3}{*}{\parbox{3cm}{\centering{integral replaced \\ by relaxation tensor}}} & \vfill identical results (not shown) \vfill \\
  \cline{2-2} \cline{4-4}
   & \vfill $S_{L,\alpha \alpha} \neq S_{L,\beta \beta};$ \vfill {\small $S_{L,\alpha \beta} = \sqrt{S_{L,\alpha \alpha}  S_{L,\beta \beta}}$} \vfill & 
   & \vfill slightly different crosspeak evolution (see Fig.~\ref{fig:plot_R2_real_correlated_fluctuations_with_and_without_integration}) \vfill
  \\
  \cline{1-2} \cline{4-4}
  \multirow{2}{*}{uncorrelated} & \multirow{2}{*}{$g_{\alpha \beta}=0$} & & \vfill different crosspeak evolution 
  (see Fig.~\ref{fig:plot_R2_real_uncorrelated_fluctuations_with_and_without_integration}) \vfill 
  \\
  \cline{3-4}
   & & \vfill $-\lambda_{ll}=\Im(\dot{g}_{ll}(t'))$ \vfill $l \in \{ \alpha,\beta \}$ \vfill & \vfill good agreement at large population times
  (see Fig.~\ref{fig:plot_R2_real_uncorrelated_fluctuations_limiting_case_with_and_without_integration}) \vfill
  \\
  \hline
\end{tabular}
\end{center}
\caption{\label{tab:overview_different_cases} Overview of the different assumptions, approximations and the main differences compared to the rigorous approach.} 
\end{table}

\newpage

\newpage

\newpage

\newpage

\newpage

\newpage

\newpage

\newpage





\providecommand{\newblock}{}

\end{document}